\documentclass[12pt]{JHEP3}
\usepackage{mathrsfs}
\usepackage{amsmath,amssymb}
\usepackage{epsfig}
\input epsf

\def\x'{\mathaccent 19 x}
\def\y'{\mathaccent 19 y}
\def\n'{\mathaccent 19 n}
\def\u'{\mathaccent 19 u}

\def\et'{\mathaccent 19 \eta}
\def\th'{\mathaccent 19 \theta}
\def\lam'{\mathaccent 19 \lambda}
\def\varet'{\mathaccent 19 \vartheta}
\def\rh'{\mathaccent 19 \rho}
\def\ph'{\mathaccent 19 \phi}
\def\xb'{\mathaccent 19 {\bar{x}}}

\def\B{\rm B}
\def\F{\rm F}

\def\l{{\lambda}}

\def\N{${\cal N}=4$ }

\def \A {{\bf{A}}}

\def\be{\begin{equation}}
\def\ee{\end{equation}}

\newcommand{\bea}{\begin{eqnarray}}
\newcommand{\eea}{\end{eqnarray}}

\def\r {\rho}
\def\a {\alpha}
\def\b {\beta}
\def\s {\sigma}
\def\pa {\partial}

\def\g {\gamma}
\def\p{\phi}
\def\la{\label}
\def\e{\epsilon}
\def\ov{\over}

\def\Vh{{\widehat V}}
\def\Uh{{\widehat U}}

\newcommand{\alg}[1]{\mathfrak{#1}}
\newcommand{\su}{\alg{su}}

\newcommand{\psu}{\alg{psu}}

\newcommand{\AdS}{{\rm  AdS}_5\times {\rm S}^5}

\def\L{\mathscr L}

\newcommand{\sfrac}[2]{{\textstyle\frac{#1}{#2}}}

\preprint{ {\tt hep-th/0512253}\\ {\tt ITP-UU-05/58} \\ {\tt
SPIN-05/41}\\
{\tt AEI-2005-186}}

\title{Green-Schwarz Strings in TsT-transformed backgrounds
}

\author{
L. F. Alday$^{a}$\footnote{e-mail: L.F.Alday@phys.uu.nl, G.Arutyunov@phys.uu.nl,
frolovs@aei.mpg.de},
G. Arutyunov$^{a}$\footnote{Also at Steklov
Mathematical Institute, Moscow }  and S. Frolov$^{b}$\footnote{
Also at
Steklov Mathematical Institute,
Moscow }
\\
\\
$^{a}$ {\it Institute for Theoretical Physics and Spinoza Institute, Utrecht University \\
~~3508 TD Utrecht, The Netherlands}\\
$^{b}$ {\it Max-Planck-Institut f\"ur Gravitationsphysik,
Albert-Einstein-Institut}\\
~~Am M\"uhlenberg 1, D-14476 Potsdam, Germany\\
}

\abstract{We consider classical strings propagating in a
background generated by a sequence of TsT transformations. We
describe a general procedure to derive the Green-Schwarz action
for strings. We show that the U(1) isometry variables of the
TsT-transformed background are related to the isometry variables
of the initial background in a universal way independent of the
details of the background. This allows us to prove that strings in
the TsT-transformed background are described by the Green-Schwarz
action for strings in the initial background subject to twisted
boundary conditions. Our construction implies that a TsT
transformation preserves integrability properties of the string
sigma model. We discuss in detail type IIB strings propagating in
the $\g_i$-deformed $\AdS$ space-time, find the twisted boundary
conditions for bosons and fermions, and use them to write down an
explicit expression for the monodromy matrix. We also discuss
string zero modes whose dynamics is governed by a fermionic
generalization of the integrable Neumann model.

}

\begin{document}

\newpage

\renewcommand{\thefootnote}{\arabic{footnote}}
\setcounter{footnote}{0}
\section{Introduction}
It is well-known that a T-duality transformation applied to a circle
which could contract to zero size produces a singular geometry from a regular one.
Recently, it was noticed in \cite{LM} that in a situation when the initial geometry
contains a two-torus a regular background may be generated by using a combination of
a T-duality transformation on one angle variable,
a shift of another isometry variable, followed by the second T-duality on the first angle.
We will refer to the chain of these transformations producing
a one-parameter deformation of the initial background as a TsT transformation.
The observation of \cite{LM} can be easily generalized to construct regular
multi-parameter deformations
of gravity backgrounds if they contain a higher-dimensional torus \cite{F}
by using a chain of TsT transformations.

A TsT transformation appears to be very useful in a search of new
less supersymmetric examples of the AdS/CFT correspondence
\cite{M}. In particular, it was successfully used in \cite{LM} to
obtain a deformation of the $\AdS$ geometry which was conjectured
to be dual to a supersymmetric marginal deformation of \N SYM
sometimes called a $\b$ deformation \cite{LEST}-\cite{BECH}.
Various aspects of the deformed gauge and string theories, and the
conjectured duality have been studied in
\cite{FRT1,deMelloKoch:2005vq}
 by
using the ideas and methods developed to test the duality between
the undeformed models \cite{BMN,FT}%-\cite{FT}
.

Strings in the more general three-parameter deformed $\AdS$
background \cite{F}, and the dual nonsupersymmetric deformation of
\N SYM have been studied in \cite{BR}-\cite{deMelloKoch:2005jg}.
It is unclear, however, if the nonsupersymmetric string background
is stable,\footnote{ It is known that the spectrum of string
theory in the TsT-transformed flat space contains tachyons
\cite{Russo}. It does not imply that string theory on the deformed
$\AdS$ is unstable because the TsT-transformed flat space is
singular at space infinity while the deformed $\AdS$ is regular
everywhere. In fact, it seems that a TsT-transformation produces a
nonsingular background only if the two-torus is of a finite size.}
and the double-trace operators are not generated in the  deformed
gauge theory, thus, breaking conformal invariance as it happens
for instance in nonsupersymmetric orbifold models \cite{DKR}.

TsT transformations have been also used to deform other
interesting string backgrounds
\cite{Gauntlett:2005jb}%-\cite{Hernandez:2005xd}
. Further related results can be found in
\cite{Spradlin:2005sv,Rashkov:2005mi}%-\cite{Bundzik:2005zg}
.

A nice property of a TsT transformation is that it can be
implemented on the string sigma model level leading to simple
relations between string coordinates of the initial and
TsT-transformed background \cite{F}. The relations have been used
to show that classical solutions of string theory equations of
motion in a deformed background are in one-to-one correspondence
with those in the initial background with twisted boundary
conditions imposed on the U(1) isometry fields parametrizing the
torus. An interesting property of the twist is that it depends on
the conserved U(1) charges of the model.

The consideration in \cite{F} was restricted to the bosonic part
of type IIB Green-Schwarz superstring action on the deformed
$\AdS$. Dealing with the Green-Schwarz superstring we face a new
problem of how to define the TsT transformation for fermionic
variables. The answer is not immediately clear, because the
operation of T-duality must include a change of the fermionic
chirality. The TsT transformation involves the angle variables
which transform under the commuting isometries of the five-sphere.
Generically, fermions of the Green-Schwarz superstring on $\AdS$
also transform under the same isometries. A key idea which allows
us to solve the problem is to redefine the original fermions in
such a way that they become neutral under the isometries in
question. After this redefinition is found we can perform the
TsT-transformations on the angle variables with fermions being
just the spectators. The very existence of such a redefinition is
non-trivial and will be established in section 3.

The aim of the current paper is to extend the discussion in
\cite{F} to the most general case of a fermionic string
propagating in an arbitrary background possessing several U(1)
isometries. We analyze a TsT transformation and show that if
fermions are neutral under the isometries then the relations are
universal and do not depend on the details of the background in
complete accord with the expectations in \cite{F}. In the case of
Green-Schwarz strings in the deformed $\AdS$ background our
consideration implies the existence of a Lax pair representation,
and, therefore, classical integrability of the model.

The plan of the paper is as follows. In section 2 we consider a
general sigma model action for fermionic strings propagating in a
curved background. We assume that the action is invariant under at
least two U(1) isometry transformations. Each U(1) transformation
is realized as a shift of an angle variables with all other
bosonic and fermionic fields being neutral under the shift. We
then perform a TsT transformation on a torus parametrized by any
two of the angles, and find a TsT-transformed action. We show that
the TsT transformation preserves the U(1) currents corresponding
to the angles, and, moreover, the TsT-transformed angles are
related to the original angles by exactly the same formulas as the
ones derived for the pure bosonic case in \cite{F} leading to the
same twisted boundary conditions for the angle variables. This
implies that strings in the TsT-transformed background are
described by the Green-Schwarz action for strings in the initial
background subject to the twisted boundary conditions. We point
out that if the original Green-Schwarz string action is integrable
then the TsT-transformed action is also integrable extending the
consideration of \cite{F} to the general case. Further we  discuss
the chains of TsT transformations applied to a background
containing a $d$-dimensional torus, and show that the most general
deformation is parametrized by a skew-symmetric $d\times
d$-dimensional matrix which determines twisted boundary conditions
for the U(1) isometry variables. The results obtained in section 2
have a partial intersection with those of \cite{Rashkov:2005mi}
where a general bosonic string background was considered.

In section 3 we apply a sequence of TsT transformations to the
Green-Schwarz superstring in $\AdS$ \cite{Metsaev:1998it} to
generate the Green-Schwarz action for nonsupersymmetric strings in
the $\g_i$-deformed $\AdS$ space-time. We explain how to redefine
the bosonic and fermionic fields so that the U(1) isometry
transformations would be realized as shifts of the angle
variables. We then use the considerations in section 2 to find the
twisted boundary conditions for bosons and fermions, and conclude
that the integrability of superstrings in $\AdS$
\cite{Bena:2003wd} implies the integrability of the fermionic
string in the $\g_i$-deformed $\AdS$ space-time. We use the Lax
pair for Green-Schwarz superstrings in $\AdS$ and the twisted
boundary conditions to derive the monodromy matrix for strings in
the $\g_i$-deformed $\AdS$. The monodromy matrix can be used to
analyze the spectrum of classical strings in  the deformed
background.

In section 4 we discuss the zero-mode part of the Green-Schwarz action
for nonsupersymmetric strings in the $\g_i$-deformed $\AdS$ space-time.
It describes a particle with fermionic degrees of freedom moving in the
deformed background. The particle action is integrable,
and generalizes the well-known Neumann model to the fermionic case. The
Lax pair for the model is induced
by the Lax pair for strings in the deformed background. Quantization of
the fermionic Neumann model
should describe the spectrum of type IIB supergravity on the nonsupersymmetric $\g_i$-deformed
background.

In Conclusion we summarize the results obtained and discuss open
problems. In appendices we collect some useful formulae.

%%%%%%%%%%%%%%%%%%%%%%%%%%%%%%%%%%%%%%%%%%%%%%%%%%%%%%%%%%%%%%%%%%%%%%%%%%%%%%%%%
\section{The $\gamma$-deformed action}
%%%%%%%%%%%%%%%%%%%%%%%%%%%%%%%%%%%%%%%%%%%%%%%%%%%%%%%%%%%%%%%%%%%%%%%%%%%%%%

We start with the following general sigma model action describing
propagation of a fermionic closed string in a background with
several U(1) isometries \bea \la{Ba1} S = -{\sqrt{\lambda}\over
2}\int\, d\tau {d\s\over 2\pi} &&\left[ \g^{\a\b}\partial_\a
\phi^i\partial_\b \phi^j\, G_{ij}^0 - \e^{\a\b}\partial_\a
\phi^i\partial_\b \phi^j\, B_{ij}^0 \right.\\\nonumber &&\left. +
2\partial_\a \phi^i\left( \g^{\a\b}{U}_{\b,i}^0
-\e^{\a\b}{V}_{\b,i}^0\right) +{\cal L}_{{\rm rest}}^0\right]\,
.~~~~~~~ \eea Here ${\sqrt{\lambda}\over 2\pi}$ is the effective
string tension which is identified with the 't Hooft coupling in
the AdS/CFT correspondence, $\e^{01}\equiv\e^{\tau\s}=1$ and
$\g^{\a\b}\equiv \sqrt{-h}\, h^{\a\b}$, where $h^{\a\b}$ is a
world-sheet metric with Minkowski signature. In the conformal
gauge $\g^{\a\b} = \mbox{diag}(-1,1)$ although in the following we
will not attempt to fix any gauge. We assume that the action is
invariant under U(1) isometry transformations geometrically
realized as shifts of the angle variables $\phi_i$, $i=1,2,\ldots
,d$. That means that the string background contains a
$d$-dimensional torus $T^d$. We show explicitly the dependence of
the action on $\p_i$, and their coupling to the background fields
$G_{ij}^0$, $B_{ij}^0$ and ${U}_{\b,i}^0,{V}_{\b,i}^0$ which
generalizes the usual coupling of bosons to the target space
metric and B-field. These background fields are independent of
$\phi^i$ but can depend on other bosonic and fermionic string
coordinates which are neutral under the U(1) isometry
transformations. By ${\cal L}_{{\rm rest}}^0$ we denote the part
of the Lagrangian which depends on these other fields of the
theory. We will see in the next section that the Green-Schwarz
action for superstrings on $\AdS$ \cite{Metsaev:1998it} can be
cast to the form (\ref{Ba1}).

The action has $d$ global symmetries
corresponding to constant shifts of $\phi's$. The corresponding
Noether currents are
\bea
J_i^{\alpha}(\phi)=-\sqrt{\lambda}\Big(\g^{\a\b}\partial_\b
\phi^j\, G_{ij}^0 - \e^{\a\b}\partial_\b \phi^j\, B_{ij}^0 +
\g^{\a\b}{U}_{\b,i}^0 -\e^{\a\b}{V}_{\b,i}^0 \Big)\,  ,
\eea
and they
are conserved, $\pa_{\a}J^{\a}_i=0$, as the consequence of the
dynamical equations.

Now we perform a TsT transformation of the angle variables. To
this end we pick up a two-torus, for instance, the one, generated
by $\phi_1$ and $\phi_2$. The TsT transformation consists in
dualizing the variable $\phi_1$ with the further shift $\phi_2\to
\phi_2+\hat{\gamma} \phi_1$ and dualizing $\phi_1$ back.
Application of the TsT transformation can be symbolically
expressed as the change of variables \bea
(\phi_1,\phi_2)\stackrel{\rm TsT}{\rightarrow}
(\tilde{\phi}_1,\tilde{\phi}_2)\, . \eea The procedure to
construct the TsT-transformed action is explained in appendices A
and B. The corresponding action can be written in the same fashion
as the original one \bea \la{Ba2} S = -{\sqrt{\lambda}\over
2}\int\, d\tau {d\s\over 2\pi} &&\left[ \g^{\a\b}\partial_\a
\tilde{\phi^i}\partial_\b \tilde{\phi}^j\, G_{ij} -
\e^{\a\b}\partial_\a \tilde{\phi}^i\partial_\b \tilde{\phi}^j\,
B_{ij} \right.\\\nonumber &&\left. + 2\partial_\a
\tilde{\phi}^i\left( \g^{\a\b}{U}_{\b,i}
-\e^{\a\b}{V}_{\b,i}\right) +{\cal L}_{{\rm rest}}\right]\,  \eea
with the new fields $G_{ij}$, etc  given in terms of the original
ones. The explicit relations are listed in appendix B. Clearly,
the new action also has the same number of symmetries related to
the constant shifts of the variables $\tilde{\phi}^i$. The
conserved Noether currents have now the form \bea
\tilde{J}_i^{\alpha}(\tilde{\phi})=-\sqrt{\lambda}\Big(\g^{\a\b}\partial_\b
\tilde{\phi}^j\, G_{ij} - \e^{\a\b}\partial_\b \tilde{\phi}^j\,
B_{ij} + \g^{\a\b}{U}_{\b,i} -\e^{\a\b}{V}_{\b,i}\Big) \, .\eea
The relation between the dual variables $\tilde{\phi}$ and the
original ones $\phi$ can be found by using the formulas from
appendices A and B, and is given by
\begin{eqnarray}\nonumber
\partial_\alpha \tilde{\phi}^1&=&\partial_\alpha \phi^1-\hat{\gamma}
\epsilon_{\alpha \beta} \gamma^{\beta \tilde{\beta}}
\partial_{\tilde{\beta}} \phi^i G_{i 2}+\hat{\gamma}
\partial_\alpha \phi^i B_{i2}-\hat{\gamma} \epsilon_{\alpha \beta}
\gamma^{\beta
\tilde{\beta}}U_{\tilde{\beta}2}-\hat{\gamma} V_{\alpha 2}\\\nonumber
\partial_\alpha \tilde{\phi}^2&=&\partial_\alpha \phi^2+\hat{\gamma}
\epsilon_{\alpha \beta} \gamma^{\beta \tilde{\beta}}
\partial_{\tilde{\beta}} \phi^i G_{i 1}-\hat{\gamma}
\partial_\alpha \phi^i B_{i 1}+\hat{\gamma} \epsilon_{\alpha \beta}
\gamma^{\beta
\tilde{\beta}}U_{\tilde{\beta}1}+\hat{\gamma} V_{\alpha 1}\\\la{relpp}
\partial_\alpha \tilde{\phi}^i&=&\partial_\alpha \phi^i\, ,\quad i\ge 3
\end{eqnarray}
Using these transformation rules, one can check that the following
relation holds \bea \la{uc}
\tilde{J}_i^{\alpha}(\tilde{\phi})=J_i^{\alpha}(\phi)\, . \eea It
shows that independently of the form of the action (\ref{Ba1}) and
the presence of fermions the TsT transformation preserves the U(1)
isometry currents corresponding to the angles $\phi_i$, thus,
generalizing and proving the considerations in \cite{F} (see, also
\cite{Rashkov:2005mi} where an arbitrary bosonic background was
analyzed).

The equality (\ref{uc}) of the original and the TsT-transformed
currents also shows that the TsT-transformation is a particular
example of the {\it B\"acklund transformations}. Indeed, in full
generality the B\"acklund transformation is defined as follows
\cite{Arutyunov:2005nk} \bea \label{BT}
\tilde{J}^{\a}-J^{\a}=\epsilon^{\a\b}\pa_{\beta}\chi \eea for some
function $\chi$. Here $J^{\a}$ and $\tilde{J}^{\a}$ correspond to
the global Noether currents computed on the original and on the
B\"acklund transformed solutions respectively. Eq.(\ref{BT})
states that the difference between two currents conserved
dynamically, the original and the B\"acklund transformed, is
proportional to the trivially conserved topological current. The
TsT-transformation simply corresponds to taking $\chi=0$. However,
in our present situation we do not require that the B\"acklund
transformations should preserve the boundary conditions for the
fundamental fields of the theory.\footnote{It would be interesting
to study the general B\"acklund transformation with a non-trivial
function $\chi$ but without imposing the same boundary conditions
on the original and transformed fields. This should lead to an
alternative proof of integrability of strings in the
$\gamma$-deformed background, in the spirit of
\cite{Arutyunov:2003rg,Arutyunov:2005nk}.}

The relation (\ref{uc}) allows one to find a relation between the
$\sigma$-derivatives of the original and transformed angles \bea
\la{relm} \tilde{\phi}'_1-\phi'_1&=&-\gamma J^{\tau}_2\, ,\qquad
\hat{\gamma} = \sqrt{\lambda}\gamma\\\nonumber
\tilde{\phi}'_2-\phi'_2&=&\gamma J^{\tau}_1 \, ,\\
\nonumber \tilde{\phi}'_i-\phi'_i&=&0\, ,\quad i\ge 3\, .
\eea
Here $J^{\tau}$
means the $\tau$-component of the conserved current. This is the
{\it same} relation as was found in the bosonic case \cite{F}.

Since we consider the closed strings on the $\gamma$-deformed
background the angles $\tilde{\phi}_i$ have the following
periodicity conditions \bea \la{upc}
\tilde{\phi}_i(2\pi)-\tilde{\phi}_i(0)=2\pi n_i\, ,~~~~~n_i\in
\mathbb{Z}\, . \eea Then integrating eqs.(\ref{relm}) we obtain
the twisted boundary conditions for the original angles $\p_1$ and
$\p_2$, and the usual periodicity conditions (\ref{upc}) for the
other $d-2$ angles \bea \label{tbcbm}
\phi_1(2\pi)-\phi_1(0)&=&2\pi (n_1 + \g J_2)\, ,\\\nonumber
\phi_2(2\pi)-\phi_2(0)&=&2\pi (n_2 - \g J_1)\, , \eea where
$$
J_i=\int_0^{2\pi} \frac{{\rm d}\sigma}{2\pi} J_i^{\tau}\,
$$
is the corresponding Noether charge. We see that the twisted
boundary conditions are universal and do not depend on the details
of the background and the presence of fermions. They depend only
on the angles involved in the TsT transformation, and the total
U(1) charges.

To understand better the meaning of the relations (\ref{uc}) and
(\ref{relm}) we notice that the time components of the U(1)
currents coincide with the momenta canonically conjugated to the
angles $\phi_i$: $J_i^{\tau} = p_i = \delta S/\delta \dot{\p}_i$.
Therefore, (\ref{uc}) and (\ref{relm}) can be written in the form
\bea \la{haf} \tilde{p_i} =p_i\ ,\qquad
\tilde{\phi}'_i=\phi'_i-\gamma_{ij}p_j\, ,\qquad i,j = 1,2,\ldots
,d\, , \eea where we take summation over $j$, and $\g_{ij}$ is
skew-symmetric, $\g_{ij} = -\g_{ji}$, with just one nonvanishing
component equal to the deformation parameter: $\g_{12} = \g$.

It is obvious from the relations (\ref{haf}) that up to the
twisted boundary conditions a TsT transformation is just a simple
linear canonical transformation of the U(1) isometry variables. It
is the twist that makes the original and TsT-transformed theories
inequivalent. It is also clear that the most general
multi-parameter TsT-transformed background obtained by applying
TsT transformations successively,  many times, each time picking
up a new torus and a new deformation parameter, is completely
characterized by the relations (\ref{haf}) with an arbitrary
skew-symmetric matrix $\g_{ij}$. Therefore, a background
containing a $d$-dimensional torus admits a $d(d-1)/2$-parameter
TsT deformation. In particular, the most general TsT-transformed
$\AdS$ background with TsT transformations applied only to the
five-sphere ${\rm S}^5$ (to preserve the isometry group of ${\rm
AdS}_5$) has three independent parameters, and, therefore, is the
one found in \cite{F}.\footnote{Let us note that a Ts...sT
transformation discussed in \cite{Rashkov:2005mi} is just a
sequence of TsT transformations applied to the tori $(\p_1,\p_i)$.
The two-parameter deformation of $\AdS$ they considered is,
therefore, a particular case of the general three-parameter
deformation.} The twisted boundary conditions for the original
angles $\p_i$ in the case of the most general deformation take the
form \bea \label{tbcbgen} \phi_i(2\pi)-\phi_i(0)&=&2\pi \left(n_i
-\nu_i\right)\, ,\qquad \nu_i = -\g_{ik}\,J_k\, . \eea Notice,
that the twists $\nu_i$ always satisfy the restriction $\nu_i\,J_i
=0$.
\medskip

In the next section we will discuss the most general
three-parameter deformation of the $\AdS$ background. For reader's
convenience below we specialize our formulae to this case.

The general three-parameter $\gamma$-deformed background is
obtained by applying the TsT transformation three
times. We express the corresponding procedure as
 \bea
(\phi_1,\phi_2,\phi_3)\stackrel{\gamma_3}{\rightarrow}
(\tilde{\phi}_1,\tilde{\phi}_2,\tilde{\phi}_3)
\stackrel{\gamma_1}{\rightarrow}
(\tilde{\tilde{\phi}}_1,\tilde{\tilde{\phi}}_2,\tilde{\tilde{\phi}}_3)
\stackrel{\gamma_2}{\rightarrow}
(\check{\phi}_1,\check{\phi}_2,\check{\phi}_3) \, .
\eea
Since
under every step the corresponding Noether currents remain the
same we can summarize relation between the angles in the following
table
\bea
\begin{array}{lll}
\tilde{\phi}'_1-\phi'_1=-\gamma_3
J^{\tau}_2\,  ~~~&~~~ \tilde{\tilde{\phi}}'_1-\tilde{\phi}'_1=0\,
~~~&~~~ \check{\phi}_1-\tilde{\tilde{\phi}}'_1=\gamma_2 J^{\tau}_3\,
\\
\tilde{\phi}'_2-\phi'_2=\gamma_3 J^{\tau}_1 \,  ~~~&~~~
\tilde{\tilde{\phi}}'_2-\tilde{\phi}'_2=-\gamma_1 J^{\tau}_3 \,
~~~&~~~ \check{\phi}'_2 -\tilde{\tilde{\phi}}'_2=0\,
\\
\tilde{\phi}'_3-\phi'_3=0\, ~~~&~~~
\tilde{\tilde{\phi}}'_3-\tilde{\phi}'_3=\gamma_1 J^{\tau}_2 \,
~~~&~~~ \check{\phi}'_3-\tilde{\tilde{\phi}}'_3=-\gamma_2
J^{\tau}_1 \,
\end{array}
 \eea
From here we straightforwardly find the relation between the
derivatives of the angles $\phi_i$ and the derivatives of
$\check{\phi}_i$, the latter being attributed to string on the
$\gamma$-deformed background: \bea \label{ra}
\check{\phi}'_i-\phi'_i=\e_{ijk}\gamma_jJ_k^{\tau}\, . \eea We see
from the formula that $\g_{ik} = -\e_{ijk}\gamma_j$. Integrating
eq.(\ref{ra}) and taking into account that
$\check{\phi}_i(2\pi)-\check{\phi}_i(0)=2\pi n_i\, , n_i\in
\mathbb{Z}$, we obtain the twisted boundary conditions for the
original angles \bea \label{tbcb} \phi_i(2\pi)-\phi_i(0)=2\pi (n_i
-\nu_i )\, ,\qquad \nu_i = \e_{ijk}\gamma_jJ_k\, .\eea

%%%%%%%%%%%%%%%%%%%%%%%%%%%%%%%%%%%%%%%%%%%%%%%%%%%%%%%%%%%%%%%%%%%%%%%%
\section{Green-Schwarz strings in $\g_i$-deformed $\AdS$}

In this section we apply TsT transformations to the Green-Schwarz
superstring in $\AdS$ \cite{Metsaev:1998it} to generate
nonsupersymmetric Green-Schwarz action for strings in the
$\g_i$-deformed $\AdS$ space-time. To this end we need to redefine
the bosonic and fermionic fields so that the U(1) isometry
transformations would be realized as shifts of the angle
variables. We then use the considerations in section 2 to find the
twisted boundary conditions for bosons and fermions, and conclude
that the integrability of superstrings in $\AdS$
\cite{Bena:2003wd} implies the integrability of the fermionic
string in the $\g_i$-deformed $\AdS$ space-time. We use the Lax
pair for Green-Schwarz superstrings in $\AdS$ and the twisted
boundary conditions to derive the monodromy matrix for strings in
the $\g_i$-deformed $\AdS$.

%%%%%%%%%%%%%%%%%%%%%%%%%%%%%%%%%%%%%%%%%%%%%%%%%%%%%%%%%%%%%%%%%%%%%%%%%
\subsection{Superstring on $\AdS$ as the coset sigma-model}
The Green-Schwarz superstring on $\AdS$ can be described as the
sigma model whose target-space is the coset \cite{Metsaev:1998it}
$$\frac{{\rm PSU}(2,2|4)}{{\rm
SO(4,1)}\times{\rm SO(5)}}\, ,$$ where ${\rm PSU}(2,2|4)$ is
supergroup of the superconformal algebra $\psu(2,2|4)$. In what
follows we will use the convention of
\cite{Alday:2005jm}\footnote{See also \cite{Alday:2005gi} and
\cite{Alday:2005jm} for the introduction into the theory of the
superalgebra $\psu(2,2|4)$. }.

Consider a group element $g$ belonging to ${\rm PSU}(2,2|4)$ and
construct the following current \bea \label{la} \A=-g^{-1}{\rm
d}g=\underbrace{\A^{(0)}+\A^{(2)}}_{\rm even
}+\underbrace{\A^{(1)}+\A^{(3)}}_{\rm odd}\, . \eea We recall that
$\psu(2,2|4)$ admits a $\mathbb{Z}_4$-grading automorphism with
respect to which it decomposes as the vector space into the direct
sum of four components: two of them are even (bosons) and two are
odd (fermions). In eq.(\ref{la}) $\A^{(0,2)}$ are bosonic
elements, and $\A^{(1,3)}$ are the fermionic ones. By construction
the current $\A$ is flat, {\it i.e.} it has the vanishing
curvature. Then the Lagrangian density for superstring on $\AdS$
can be written in the form \cite{Metsaev:1998it,Roiban} \bea
\label{sLag} \L =-\sfrac{1}{2}\sqrt{\lambda}\,{\rm
str}\Big(\gamma^{\a\b}\A^{(2)}_{\a}\A^{(2)}_{\b}+\kappa
\epsilon^{\a\beta}\A^{(1)}_{\a}\A^{(3)}_{\beta}\Big)\, , \eea
which is the sum of the kinetic and the Wess-Zumino terms, and
$\kappa$-symmetry requires $\kappa=\pm 1$.

\medskip
The next step is related to an explicit choice of the coset
representative $g$. As was shown in \cite{Alday:2005jm} a
convenient parametrization is provided by choosing  \bea
\label{param} g=g(\theta)g(z). \eea Here $g(\theta)\equiv
\exp(\theta)$, where $\theta$ is an odd element of $\psu(2,2|4)$
which comprises 32 fermionic degrees of freedom. The element
$g(z)$ belongs to SU(2,2)$\times$ SU(4). The coordinates $z\equiv
(x_a,y_a)$ with $a=1,\ldots, 5$ parametrize the five-sphere and
${\rm AdS}_5$ respectively.

\medskip

\noindent With parametrization (\ref{param}) we get for the flat
current the following representation
\bea
\label{A}
\A = -
g^{-1}dg = - g^{-1}(z) g^{-1}(\theta)dg(\theta)g(z) - g^{-1}(z)
dg(z)\, .
\eea
Since
$$g(\theta) = \cosh\theta + \sinh\theta\, ,\quad\quad
g^{-1}(\theta) = \cosh\theta - \sinh\theta
$$
we see that
\bea
\label{gdg}
g^{-1}(\theta)dg(\theta) ={\rm F} +
{\rm B}\, ,
\eea
where
\bea \nonumber {\rm B}&\equiv
&\cosh\theta\,
d\cosh\theta -\sinh\theta\, d\sinh\theta \, ,\\
 {\rm F}&\equiv &  \cosh\theta\,
d\sinh\theta -\sinh\theta\, d\cosh\theta \eea are the even
(boson) and odd (fermion) elements respectively. Thus, the even
component of $\A$ is
\bea \label{evenA}
\A_{\rm even} = -
g^{-1}(z) {\rm B} g(z) - g^{-1}(z) dg(z)\, ,
\eea
while the odd
component is
\bea \label{oddA} \A_{\rm odd} = - g^{-1}(z) {\rm F}
g(z) \, .
\eea
It is interesting to note that for such a
parametrization of the coset the even component of the flat
current is a gauge transform of the even element $\B$, while the
odd component is conjugate to $\F$ with the bosonic matrix $g(z)$.

\medskip
To write down the final Lagrangian we have to find the projections
$\A^{(i)}$. This can be easily done by using an explicit
representation for the action of the $\mathbb{Z}_4$-grading
automorphism and we refer the reader to \cite{Alday:2005jm} for
the corresponding discussion. To present further results we
introduce two $8\times 8$ matrices
$$
K_8 = \left(
\begin{array}{cc}
  K & 0  \\
  0 & K
\end{array} \right)\, ,\qquad \widetilde{K}_8 = \left(
\begin{array}{cc}
  K & 0  \\
  0 & -K
\end{array} \right)\, ,
$$
where $K$ is a $4\times 4$ matrix obeying the condition
$K^2=-\mathbb{I}$. These matrices are used to define
$$
{\rm G} = g(z) K_8 g(z)^t \equiv \left(
\begin{array}{cc}
  g_a & 0  \\
  0   & g_s
\end{array} \right)\, ,\quad \tilde{\rm G} = g(z) \widetilde{K}_8 g(z)^t
\equiv \left(
\begin{array}{cc}
  g_a & 0  \\
  0 & -g_s
\end{array} \right)\, .
$$
As was discussed in \cite{Alday:2005jm}, the $4\times 4$ matrices
$g_a\in {\rm SU(2,2)}$ and $g_s\in {\rm SU(4)}$ provide another
parametrization of the five-sphere and the ${\rm AdS}$ space. On
coordinates $z$ the global symmetry algebra $\psu(2,2|4)$ is
realized non-linearly. In opposite, $g_a$ and $g_s$ carry a linear
representation of the superconformal algebra. Such realization of
symmetries makes an identification of string states with operators
of the dual gauge theory more transparent. We further find
\bea \nonumber 2\, \A^{(0)} &=& \A_{\rm even} + K_8 \A_{\rm
even}^t K_8 =-2g^{-1}dg -g^{-1}\left( {\rm B} - {\rm G} {\rm B}^t
{\rm G}^{-1}
- d{\rm G} {\rm G}^{-1}\right) g\, ,~~~~~~~~~\\
\nonumber 2\, \A^{(2)} &=& \A_{\rm even} - K_8 \A_{\rm even}^t K_8
=-g^{-1}\left( {\rm B} + {\rm G} {\rm B}^t {\rm G}^{-1}
+ d{\rm G} {\rm G}^{-1}\right) g\, ,~~~~~~~~~\\
\nonumber 2\, \A^{(1)} &=& \A_{\rm odd} + i \widetilde{K}_8
\A_{\rm odd}^t K_8 =
-g^{-1}\left( {\rm F} - i \tilde{\rm G} \F^t {\rm G}^{-1} \right) g\, ,~~~~~~~~~\\
\la{tranA} 2\, \A^{(3)} &=& \A_{\rm odd} - i \widetilde{K}_8
\A_{\rm odd}^t K_8 = -g^{-1}\left( {\rm F} + i \tilde{\rm G} {\rm
F}^t {\rm G }^{-1} \right) g\, .~~~~~~~~~ \eea Substituting these
projections into the string Lagrangian (\ref{sLag}) we
obtain\footnote{For convenience we rescaled the whole Lagrangian
by the factor of $4$.}
\bea \nonumber
\L=&-&\frac{1}{2}\sqrt{\lambda}\,{\rm
str}\left[
\gamma^{\a\b}(\B_{\a}+G\B_{\a}^tG^{-1}+\pa_{\a}G G^{-1})
(\B_{\b}+G\B_{\b}^tG^{-1}+\pa_{\b}G G^{-1})\right.\\
\nonumber &+&\left.\kappa\epsilon^{\a\b}({\rm
F}_{\a}-i\tilde{\rm G}{\rm F}_{\a}^t{\rm G}^{-1})({\rm
F}_{\b}+i\tilde{\rm G}{\rm F}_{\b}^t{\rm G}^{-1})\right]\, .
\eea
By using the cyclic property of the supertrace the Wess-Zumino
term can be further simplified and we get
\bea \nonumber
\L=&-&\frac{1}{2}\sqrt{\lambda}\,{\rm
str}\left[\gamma^{\a\b}
(\B_{\a}+G\B_{\a}^tG^{-1}+\pa_{\a}G
G^{-1})(\B_{\b}+G\B_{\b}^tG^{-1}+\pa_{\b}G
G^{-1})\right.\\\la{Lor}
&+&\left. 2i\kappa\epsilon^{\a\b}{\rm F}_{\a}\tilde{\rm G}{\rm
F}_{\b}^t{\rm G}^{-1}\right]\, .
\eea The nice feature of this
Lagrangian is that it depends only on fields which carry linear
representation of the superconformal group. In particular, we have
three linearly realized U(1) isometries which are used
to construct the Green-Schwarz superstring on the
$\gamma$-deformed background.

\medskip

With a certain choice of the matrix $K$ the matrix $g_s$
parametrizing ${\rm S}^5$ can be written as follows (see, {\it
e.g.}
%the
%second paper of
\cite{AFRT}):
\bea \label{matr}
g_s=\left(\begin{array}{cccc}
0 & u_3 & u_1 & u_2 \\
-u_3 & 0 & u_2^* & -u_1^* \\
-u_1 & -u_2^* & 0 & u_3^* \\
-u_2 & u_1^* & -u_3^* & 0
\end{array}
\right) \, ,
\eea
This is the unitary matrix $g_s^\dagger g_s
=\mathbb{I}$ provided the three complex coordinates $u_i$ obey the
constraint $|u_1|^2+|u_2|^2+|u_3|^2 = 1$. A similar
parametrization of the ${\rm AdS}_5$ space is given by
\bea
\label{matra}
g_{a}=\left(\begin{array}{cccc}
0 & v_3 & v_1 & v_2 \\
-v_3 & 0 & -v_2^* &v_1^* \\
-v_1 & v_2^* & 0 & v_3^* \\
-v_2 & -v_1^* & -v_3^* & 0
\end{array}
\right)\, .
\eea
Here $g_a\in {\rm SU}(2,2)$, {\it i.e.} it obeys
$g_{a}^{\dagger}Eg_{a}=E$ with $E={\rm diag}(1,1,-1,-1)$ provided
the complex numbers $v_i$ satisfy the constraint:
$|v_1|^2+|v_2|^2-|v_3|^2=-1$.

%%%%%%%%%%%%%%%%%%%%%%%%%%%%%%%%%%%%%%%%%%%%%%%%%%%%%%%%%%%%%%%%%%%%%%%%%%%%%%%%%
\subsection{Fermions twisting}
The original fermions appearing in the Lagrangian (\ref{Lor})
transform under the commuting
isometries of the five-sphere.
To apply the consideration in section 2 to
Green-Schwarz superstrings in $\AdS$ we need to
redefine these fermions
in such a way that they become
neutral under the isometries in question.
After this redefinition
is found we can perform the TsT-transformations on the angle
variables with fermions being just the spectators, and use the general
formulas derived in section 2. The twisted boundary conditions (\ref{tbcb}) for the
original angles of $\AdS$ then induce twisted boundary conditions for
the original charged fermions of $\AdS$.

\medskip

Let us explore in more detail the invariance of the Lagrangian
under the abelian subalgebra of the superconformal group. In full
generality the bosonic symmetry algebra ${\rm SO(4,2)}\times {\rm
SO(6) }$ has six Cartan generators: three for ${\rm SO(4,2)}$ and
three for ${\rm SO(6)}$. If we introduce the polar representation
$$
u_i=r_i\, e^{i\,\p_i}\, ,  ~~~~~~~~~~v_i= \r_i\, e^{i\,\psi_i}\, ,
$$
with $r_i,\rho_i$ being real, then the six commuting isometries
are realized as constant shifts of the angle variables
$$
\phi\to \phi+\epsilon \, , ~~~~~~~\psi \to \psi+\epsilon \, .
$$
Remarkably, it turns out that the matrices $g_s$ and $g_a$ enjoy
the following factorization property \cite{F} (see also
\cite{AFRT}) \bea \label{factform}
g_s(r,\p)&=&M(\p)\, \hat{g}_s(r)\, M(\p)\, ,\\
g_a(\r,\psi)&=&M(\psi)\, \hat{g}_a(\r)\, M(\psi)\, ,\eea where \bea
\hat{g}_s(r) = \left(\begin{array}{cccc}
0 & r_3 & r_1 & r_2 \\
-r_3 & 0 & r_2 & -r_1 \\
-r_1 & -r_2 & 0 & -r_3 \\
-r_2 & r_1 & r_3 & 0
\end{array}
\right)\, , ~~~~~~ \hat{g}_a(\r) = \left(\begin{array}{cccc}
0 & \r_3 & \r_1 & \r_2 \\
-\r_3 & 0 & \r_2 & -\r_1 \\
-\r_1 & -\r_2 & 0 & \r_3 \\
-\r_2 & \r_1 & -\r_3 & 0
\end{array}
\right)\, .
\eea
Here also $M(\p)=e^{\frac{i}{2}\Phi(\phi)}$,
where $\Phi(\phi)={\rm diag}(\Phi_1,\ldots,\Phi_4)$ with
\bea
\nonumber \Phi_1&=&\p_1 +
\p_2 + \p_3 \\
\nonumber \Phi_2&=&-\p_1-\p_2+\p_3 \\
\nonumber \Phi_3&=&\p_1-\p_2-\p_3 \\
\label{Phi} \Phi_4&=&-\p_1+\p_2-\p_3
 \eea
The simplest way to see that all fermions are charged under the
six commuting isometries is to notice that any fermionic term in the Lagrangian (\ref{Lor})
explicitly depends on all the angle variable $\p_i$ and $\psi_i$.
To find the fermion redefinition that makes them neutral
we represent the odd matrix $\theta$ as
\bea
\theta=\left(\begin{array}{cc} 0 & X\\ Y & 0\end{array}\right)
\eea
Then it is clear that to uncharge the fermions under all U(1)'s we have to
make the following rescaling
\bea
X=M(\psi_i)\hat{X}M(\phi_i)^{-1}\\
Y=M(\phi_i)\hat{Y}M(\psi_i)^{-1}
\eea
This leads to the following
transformation formula
\bea
g(\theta)= \left(\begin{array}{cc}
M(\psi_i) & 0\\ 0 & M(\phi_i)\end{array}\right)g(\hat{\theta})
\left(\begin{array}{cc} M(\psi_i)^{-1} & 0\\ 0 &
M(\phi_i)^{-1}\end{array}\right)\, ,
\eea
where the fermions
$\hat{\theta}$ are uncharged under all U(1)s.

In what follows we restrict our attention to TsT transformations
applied to the five-sphere, and, therefore, we do not need to make
fermions neutral under the isometries of ${\rm AdS}_5$. The
corresponding redefinition of fermions simplifies and takes the
following form \bea\la{fermtr}
&&X=\hat{X}M(\phi_i)^{-1}\, ,\quad  Y=M(\phi_i)\hat{Y} \\
&&g(\theta)= \left(\begin{array}{cc} 1 & 0\\ 0 &
M(\phi_i)\end{array}\right)g(\hat{\theta}) \left(\begin{array}{cc}
1 & 0\\ 0 & M(\phi_i)^{-1}\end{array}\right)\, . \eea Let us
mention, however, that the fermions do have to be neutral under
some isometries of ${\rm AdS}_5$, in particular, shifts of the
global AdS time $t\equiv \psi_3$, if one wants to impose the
uniform light-cone gauge that was recently used to solve the
$\su(1|1)$ sector of superstrings in $\AdS$
\cite{Arutyunov:2005hd}.

Now, to determine the twisted boundary conditions for fermions we
just need to take into account that the redefined neutral fermions
do not transform under the TsT transformations. Therefore, the
original charged fermions in $\AdS$ satisfy twisted boundary
conditions which can be easily found by using  (\ref{fermtr}), and
the twisted boundary conditions (\ref{tbcb}) for the angles
$\phi_i$: \bea \la{twistfer} &&X(2\pi)=X(0)e^{i\pi\Lambda}\,
,\quad
Y(2\pi)=e^{-i\pi\Lambda}Y(0) \, ,\\
&&g(\theta)(2\pi)= \left(\begin{array}{cc}
1 & 0\\ 0 & e^{-i\pi\Lambda}\end{array}\right)g(\theta)(0)
\left(\begin{array}{cc} 1 & 0\\ 0 &e^{i\pi\Lambda}
\end{array}\right)\, ,
\eea where $\Lambda$ is the following diagonal matrix
$\Lambda={\rm diag}(\Lambda_1,\ldots,\Lambda_4)$ with
\bea\nonumber
\Lambda_1&=&\gamma_1(J_2-J_3)+\gamma_2(J_3-J_1)+\gamma_3(J_1-J_2)=\nu_1+\nu_2+\nu_3\\
 \nonumber
\Lambda_2&=&\gamma_1(J_2+J_3)-\gamma_2(J_1+J_3)-\gamma_3(J_1-J_2)=-\nu_1-\nu_2+\nu_3\\
\nonumber
\Lambda_3&=&-\gamma_1(J_2-J_3)+\gamma_2(J_1+J_3)-\gamma_3(J_1+J_2)=\nu_1-\nu_2-\nu_3\\
 \la{Lam}
\Lambda_4&=&-\gamma_1(J_2+J_3)-\gamma_2(J_3-J_1)+\gamma_3(J_1+J_2)=-\nu_1+\nu_2-\nu_3
\eea Obviously, the four variables $\Lambda_k$ depend on three
$\nu_i$'s precisely in the same fashion as $\Phi_k$ depend on
$\phi_i$'s, c.f. eqs.(\ref{Phi}). The formulas (\ref{tbcb}) and
(\ref{twistfer}) allow us to analyze strings in the deformed
background by using twisted strings in $\AdS$.

%%%%%%%%%%%%%%%%%%%%%%%%%%%%%%%%%%%%%%%%%%%%%%%%%%%%%%%%%%%%%%%%%%%%%%%%%%
\subsection{Lax Pair and Monodromy Matrix}
As was discussed in detail in \cite{F}, the relations (\ref{relpp}) can be used
to find a local periodic Lax pair for strings
in a TsT-transformed background if an isometry invariant
Lax pair for strings in the initial background is known. The twisted boundary conditions
(\ref{tbcb}) then can be used to get a simple expression for the TsT-transformed monodromy
matrix in terms of the initial monodromy matrix and the twist matrix.

We begin by recalling the structure of the Lax pair found in \cite{Bena:2003wd}. It is based on
the two-dimensional Lax connection $\mathscr{L}$ with components
\bea
\label{wL}
\mathscr{L}_{\a}=\ell_0 \A_{\a}^{(0)}+\ell_1 \A_{\a}^{(2)}
+\ell_2\gamma_{\a\b}\epsilon^{\beta\rho}\A_{\rho}^{(2)} +\ell_3
Q_{\a}^++\ell_4 Q_{\a}^-\, ,
\eea
where $\ell_i$
are functions of a spectral parameter, and
$Q^{\pm}=\A^{(1)}\pm \A^{(3)}$.
The zero curvature condition for the
connection $\mathscr{L}$,
\bea
\label{zc}
\pa_{\a}\mathscr{L}_{\beta}-\pa_{\b}\mathscr{L}_{\a}-
[\mathscr{L}_{\a},\mathscr{L}_{\b}]=0 \, ,
\eea
follows from the
dynamical equations and the flatness of $\A_{\a}$ if
$\ell_i$ are chosen in the form
\bea\nonumber
%\label{fp1}
\ell_0=1, ~~~~\ell_1=\frac{1+x^2}{1-x^2}\, ,~~~~\ell_2&=&s_1\frac{2x}{1-x^2}\, ,
~~~~\ell_3=s_2\frac{1}{\sqrt{1-x^2}}\,
,~~~~\ell_4=s_3\frac{x}{\sqrt{1-x^2}} \, , \eea
where $x$ is the spectral parameter, and the constants $s_i$ satisfy
\bea \nonumber
s_2^2=s_3^2&=&1\\\nonumber
s_1+\kappa s_2s_3&=&0\, . \eea
Thus, for every
choice of $\kappa$ we have four different solutions for $\ell_i$
specified by the choice of $s_2=\pm 1$ and $s_3=\pm 1$.
By using eqs.(\ref{A}) for $\A_{\a}$, the Lax connection (\ref{wL}) can be
explicitly realized in terms of $8\times 8$ supermatrices from the
Lie algebra $\su(2,2|4)$. However,
as was explained
in \cite{Alday:2005gi}, in the algebra $\su(2,2|4)$ the
curvature (\ref{zc}) of $\L_{\a}$ is not exactly zero, rather it
is proportional to the identity matrix (anomaly) with a
coefficient depending on fermionic variables. However,
since $\psu(2,2|4)$ is the factor-algebra of $\su(2,2|4)$ over its
central element proportional to the identity matrix,
the curvature is regarded to be zero
\cite{Alday:2005gi,Beisert:2005bm} in the algebra $\psu(2,2|4)$.

The Lax connection (\ref{wL}) cannot be used to derive a Lax pair for strings in
the deformed background because  $\A_{\a}$ explicitly depends on $\phi_i$, and, therefore,
$\L_{\a}$ is not isometry invariant. To get a proper Lax connection we need to make
a gauge transformation of $\L_{\a}$ similar to the one used in \cite{F} for the bosonic case.

The necessary gauge transformation can be found in two steps.
First, we use the group element $g$ and formulas (\ref{tranA}) to
derive a Lax connection $\widetilde{\L}_{\a}$ which depends only
on the coset element $G$. The transformed Lax connection still has
an explicit dependence on the angles $\p_i$, but it can be easily
gauged away by using the factorization property (\ref{factform})
of $G$, and making the fermions neutral under the U(1) isometries
of ${\rm S}^5$ by using (\ref{fermtr}). The resulting gauge
transformation that converts the Lax connection  (\ref{wL}) to an
isometry invariant form, therefore, is \bea \la{gaugetran}
h=M^{-1} g\, ,\quad \partial_\a - \L_{\a}\rightarrow
\partial_a - \hat{\L}_\a = M^{-1}\left( \partial_\a -
\widetilde{\L}_{\a}\right) M = h\left(
\partial_\a - \L_{\a}\right) h^{-1}\, , \eea where
$\widetilde{\L}_{\a} = g\L_{\a} g^{-1}+\partial_\a g g^{-1}$ can
be easily found by using (\ref{tranA}) and (\ref{wL}), and the 8
by 8 matrix $M$ is
$$
M=\left(\begin{array}{cc}
1 & 0\\ 0 & M(\phi_i)\end{array}\right)\, .
$$
The Lax connection $\hat{\L}_\a$ depends only on the derivatives of $\p_i$,
and, as was explained in \cite{F}, to get a Lax connection for strings in the deformed
$\AdS$ all one needs to do is to express $\pa_\a\p_i$ in terms of  $\pa_\a\tilde{\p}_i$
by using the relations  (\ref{relpp}).
The resulting expression for the Lax connection $\hat{\L}_\a$ is rather complicated,
and it is difficult to write down its explicit form.

The gauged transformed Lax connection $\hat{\L}_\a$ is, obviously,
flat, and is invariant under the U(1) isometries, and is periodic
in $\s$. It can be used to compute the monodromy matrix T$(x)$
which is defined as the path-ordered exponential of the spatial
component $\hat{\L}_{\sigma}(x)$ of the Lax connection \cite{FTa}
\bea \la{T} {\rm T}(x)={\cal P}\exp\int_0^{2\pi}{\rm
d}\s~\hat{\L}_{\sigma}(x) \, , \eea The key property of the
monodromy matrix is the time conservation of all its spectral
invariants. In particular, any eigenvalue of ${\rm T}(x)$,
$\exp(ip_k(x))$ where $p_k(x)$ is called a quasi-momentum,
generates an infinite set of integrals of motion.

In the context of the AdS/CFT correspondence the monodromy matrix of
the Lax connection $\L_\a$ of superstrings in $\AdS$ was used in \cite{KMMZ,Beisert:2005bm}
to derive finite-gap integral equations which describe the spectrum of
classical spinning strings in the scaling limit of \cite{FT}.

The derivation of the equations requires a careful analysis of various
asymptotic properties of the monodromy matrix T$(x)$ and
the quasi-momenta $p(x)$
at small and large values of the spectral parameter $x$.
An important distinction of $\hat{\L}_\a$ from $\L_\a$ is that it does not vanish
at large values of $x$, and that makes more difficult to
study the large $x$ asymptotic properties of the monodromy matrix.

To analyze the asymptotics it is more convenient to use the
nonlocal and nonperiodic Lax connection $\widetilde{\L}_\a$
explicitly depending on the angles $\p_i$ which satisfy the
twisted boundary conditions (\ref{tbcb}). In terms of the Lax
connection the monodromy matrix T$(x)$ takes the form
\bea\la{mon2} {\rm T}(x) = M^{-1}(2\pi)\cdot {\cal
P}\exp\int_0^{2\pi}{\rm d}\s~ \widetilde{\L}_{\sigma}(x)\cdot
M(0)\, . \eea It is clear that the monodromy matrix is not similar
to the path-ordered exponential of the Lax connection
$\widetilde{\L}_{\a}$ because the matrix $M$ is not periodic.

The quasi-momenta $p_k$ can be expressed through eigenvalues of
\bea\la{mont} \widetilde{{\rm T}}(x) = M(0)M^{-1}(2\pi)\cdot {\cal
P}\exp\int_0^{2\pi}{\rm d}\s~ \widetilde{\L}_{\sigma}(x)\, . \eea
It is not difficult to check that
$$
M(0)M^{-1}(2\pi) = \left(\begin{array}{cc} 1 & 0\\ 0 &e^{i\pi\Lambda}
\end{array}\right)\, ,
$$
where $\Lambda$ is given in (\ref{Lam}).

It would be interesting to analyze the properties of the monodromy
matrix and derive finite-gap integral  equations for the  deformed
model analogous to those derived for strings in $\AdS$ in
\cite{KMMZ,Beisert:2005bm}. It was done for the simplest $\su(2)$
sector in \cite{FRT1}.

\section{Spinning particle and Neumann model}
In section 3 we established equivalence between strings on the
$\gamma_i$-deformed background and strings on $\AdS$ with twisted
boundary conditions. This equivalence allows one to construct an
action for the ``$\gamma_i$-deformed'' spinning particle. Further
quantization of this action should lead to determination of the
spectrum of IIB supergravity compactified on the corresponding
(generically non-supersymmetric) background.

A spinning particle is the string zero mode. To obtain the
spinning particle in the $\gamma$-deformed background we have to
assume that all the embedding fields describing this background
depend on the world-sheet time $\tau$ only. Correspondingly, from
the point of view of the string on $\AdS$, this means that the
embedding bosonic fields must have the following
$\tau,\sigma$-dependence \bea \label{btbc}
u_i=r_i(\tau)e^{i\phi_i(\tau)-i\sigma\nu_i }\, ,\qquad
v_i=\rho_i(\tau)e^{i\psi_i(\tau)} \, . \eea Here $\phi_i(\tau)$
and $\psi_i(\tau)$ are the time-dependent phases and the
$\sigma$-dependence of $u_i$ reflects the twisted boundary
conditions. For the matrix ${\rm G}$ (and similar for $\tilde{\rm
G}$) this implies the following structure \bea
 {\rm
G}(\tau,\sigma)=
 \left(\begin{array}{cc} 1 ~&~ 0 \\ 0 ~&~ e^{-\frac{i}{2}\Lambda \sigma}  \end{array}\right)
{\rm G}(\tau) \left(\begin{array}{cc} 1 ~&~ 0 \\ 0 ~&~
e^{-\frac{i}{2}\Lambda \sigma}  \end{array}\right)\, . \eea The
zero modes of the fermionic fields are described in an analogous
manner \bea X(\tau,\sigma)=X(\tau)e^{\frac{i}{2}\Lambda \sigma }\,
,~~~~~~~~ Y(\tau,\sigma)=e^{-\frac{i}{2}\Lambda \sigma}Y(\tau) \,
, \eea which is equivalent to \bea \theta(\tau,\sigma)=
 \left(\begin{array}{cc} 1 ~&~ 0 \\ 0 ~&~ e^{-\frac{i}{2}\Lambda \sigma}  \end{array}\right)
\theta(\tau) \left(\begin{array}{cc} 1 ~&~ 0 \\ 0 ~&~
e^{\frac{i}{2}\Lambda \sigma}  \end{array}\right)\, .
\eea
Upon
substituting these formulae into the general string action
(\ref{Lor}) one can see that the $\sigma$-dependence cancels out
leaving behind the dependence on the deformation parameters
$\gamma_i$. As the result we obtain an action for the spinning
particle in the $\gamma$-deformed background. Since the
corresponding bosonic action is known \cite{FRT2} to be the same as the action
for the so-called Neumann-Rosochatius (NR) integrable model, we
therefore obtain the fermionic generalization of the NR model.

 If we restrict for the moment our
attention to the purely bosonic case and introduce the diagonal
metric $\eta={\rm diag}(1,1,-1)$ we find the following action \bea
\nonumber \L_{\rm
bos}&=&-2\sqrt{\lambda}\gamma^{\tau\tau}\Big(\sum_{i=1}^3
\dot{r}_i^2+{r}_i^2\dot{\phi}^2+\sum_{ij}^3\eta_{ij}\dot{\rho}_i\dot{\rho}_j
+\eta_{ij}\rho_i^2\dot{\psi}_j^2\Big)
+\frac{2\sqrt{\lambda}}{\gamma^{\tau\tau}
}\sum_{i=1}^3\nu_i^2r_i^2  \\
&-&2\sqrt{\lambda}\frac{\gamma^{\tau\sigma}}{\gamma^{\tau\tau}
}\sum_{i=1}^3\nu_ir_i^2(\nu_i\gamma^{\tau\sigma}-2\gamma^{\tau\tau}\dot{\phi}_i)\,
. \eea As usual components of the world-sheet metric are
non-dynamical and play the role of the Lagrangian multipliers. In
particular, equation of motion for $\gamma^{\tau\sigma}$ is
equivalent to the following Virasoro constraint \bea\label{pVir}
\sum_{i=1}^3\nu_ir_i^2(\gamma^{\tau\sigma}\nu_i-\gamma^{\tau\tau}\dot{\phi}_i)=0\,
. \eea

Assume now that our particle rotates both in five-sphere with
angular momenta $J_i$ and also in ${\rm AdS}_5$ with spins
$S_i$.\footnote{The spin $S_3$ coincides with the space-time
energy of the particle.} Fixing $J_i$ and $S_i$ we can integrate
all the time-dependent phases $\phi_i(\tau)$, $\psi_i(\tau)$ out
by using their equations of motion. Indeed, we have
$$
\dot{\psi}_i=-\frac{\eta_{ij}S_j}{4\sqrt{\lambda}\gamma^{\tau\tau}\rho_i^2}\,
, \qquad\quad
\dot{\phi}_i=-\frac{J_i}{4\sqrt{\lambda}\gamma^{\tau\tau}r_i^2}+
\nu_i\frac{\gamma^{\tau\sigma}}{\gamma^{\tau\tau}}\, .
%\dot{\psi}_2=-\frac{S_2}{4\sqrt{\lambda}\gamma^{\tau\tau}\rho_2^2}\,
%, \qquad
%\dot{\psi}_3=\frac{S_3}{4\sqrt{\lambda}\gamma^{\tau\tau}\rho_3^2}\,
$$
 Upon substituting this solution for all six angle variables
we obtain the following bosonic action
\bea
\L_{\rm bos}&=&
\nonumber
-2\sqrt{\lambda}\gamma^{\tau\tau}(\dot{r}_1^2+\dot{r}_2^2+\dot{r}_3^2+
\dot{\rho}_1^2+\dot{\rho}_2^2-\dot{\rho}_3^2)\\
\label{bsp} &-&
\frac{1}{8\sqrt{\lambda}\gamma^{\tau\tau}}\Big(\frac{J_1^2}{r_1^2}+
\frac{J_2^2}{r_2^2}+\frac{J_3^2}{r_3^2}
+\frac{S_1^2}{\rho_1^2}+\frac{S_2^2}{\rho_2^2}-\frac{S_3^2}{\rho_3^2}-
16\lambda\sum_{i=1}^3\nu_i^2r_i^2\Big)\,
.
\eea
This is an action of the integrable NR system written in an
arbitrary world-line metric density $\gamma^{\tau\tau}$. Notice
that the second independent component of the metric
$\gamma^{\tau\sigma}$ cancels out from the action. On the other
hand, the Virasoro constraint (\ref{pVir}) reduces to \bea
\label{nuJ} \sum_i\nu_iJ_i=0 \eea with the general solution
$\nu_i=\epsilon_{ijk}\gamma_jJ_k$. Thus, if we would start with
arbitrary parameters $\nu_i$ defining the twisted boundary
conditions (\ref{btbc}), compatibility of the dynamics with the
Virasoro constraints would require that
$\nu_i=\epsilon_{ijk}\gamma_jJ_k$. This provides a new interesting
interpretation of the equation (\ref{nuJ}).

In the general fermionic case it is also possible to integrate out
the angle variables provided the fermions are redefined to be
neutral under all U(1) isometries. This redefinition has been
already discussed in the previous section and therefore we will
not repeat it here. Introducing the 16 complex uncharged fermions
$\theta=\{\theta_{\a}\}_{\a=1,\ldots, 16}$ we integrate out the
angle variables and obtain the fermionic generalization of the NR
model. Due to the complexity of the explicit answer, below we
indicate  the structure of the quadratic fermionic action only. It
reads \bea \nonumber \L_{\rm
2ferm}&=&\sqrt{\lambda}\gamma^{\tau\tau}\Big(\epsilon_{ijk}r_j
\dot{r}_k(\theta^*\Upsilon^i_r\dot{\theta}
-\dot{\theta}^*\Upsilon^i_r\theta)+\epsilon_{ijk}\rho_j
\dot{\rho}_k(\theta^*\Upsilon^i_{\rho}\dot{\theta}
-\dot{\theta}^*\Upsilon^i_{\rho}\theta)\Big)\\
\nonumber &+&\sqrt{\lambda}\kappa\Big(
r_i\rho_j\theta\Omega^{ij}\dot{\theta}+r_i\rho_j\theta^*\Omega^{ij}\dot{\theta}^*\Big)+
\frac{\sqrt{\lambda}}{\gamma^{\tau\tau}}\, r_ir_j\theta^*
\Sigma^{ij}
\theta\\
\label{fa} &+&\frac{1}{8\sqrt{\lambda}\gamma^{\tau\tau}}
\theta^*(T_1+T_2) \theta  \, .\eea Here the matrices
$\Upsilon_{r,\rho}^i$ are constant $16\times 16 $ anti-symmetric
matrices.  Matrices $\Omega^{ij}$ and $\Sigma^{ij}$ are symmetric
under $i\leftrightarrow j$ and they depend on the deformation
parameters $\nu_i$; they vanish if $\nu_i\to 0$. The explicit
formulas for the matrices $T_1$ and $T_2$ can be found in appendix
C. These matrices depend non-trivially on all the spins as well as
on coordinates $r_i$ and $\rho_i$ but they are independent of
$\nu_i$.

Since we have not attempted  to fix the $\kappa$-symmetry the
action (\ref{fa}) still depends on 32 fermionic degrees of freedom
and the kinetic term for fermions appears to be degenerate
reflecting thereby the presence of the $\kappa$-symmetry. Finally
we note that the fermionic NR model remains integrable, because
the Lax connection for the string on the $\gamma$-deformed
background admits further reduction to zero modes. It would be
very interesting to further investigate the integrable properties
of the fermionic NR model and ultimately to quantize it.

%%%%%%%%%%%%%%%%%%%%%%%%%%%%%%%%%%%%%%%%%%%%%%%%%%%%%
\section{Conclusion}
%%%%%%%%%%%%%%%%%%%%%%%%%%%%%%%%%%%%%%%%%%%%%%%%%%%%%%%%
In this paper we have discussed
classical strings propagating in a background
obtained from an arbitrary string theory background by a sequence of
TsT transformations.

Assuming that the initial background is invariant under $d$ U(1)
isometries, we have described a procedure to derive the most
general $d(d-1)$ parameter deformation of the background, and the
Green-Schwarz action governing the dynamics of the strings.

We have shown that angle variables of a TsT-transformed background
are related to angle variables of the initial background in a
universal way independent of the particular form of the background
metric and other fields. This has allowed us to prove that strings
in the TsT-transformed background are described by the
Green-Schwarz action for strings in the initial background with
bosonic and fermionic fields subject to twisted boundary
conditions. Due to this relation for many purposes it is not
necessary to know the explicit Green-Schwarz action for strings in
a TsT-transformed background. These strings can be analyze by
mapping them to twisted strings in the initial background. We have
stressed that our construction implies that a TsT transformation
preserves integrability properties of string sigma model.

We have discussed in detail type IIB strings propagating in
$\g_i$-deformed $\AdS$ space-time and found the twisted boundary conditions
for bosons and fermions. We then have used a known Lax pair for superstrings in $\AdS$, and
the relation between the angles to derive a local and periodic Lax representation
for the $\g_i$-deformed  model. The existence of the Lax pair implies the integrability
of the fermionic string sigma model on the deformed background generalizing the construction
of \cite{F}. The twisted boundary conditions for string coordinates have been used
to write down an explicit expression for the TsT-transformed monodromy matrix
in terms of the $\AdS$ monodromy matrix, and the twist matrix.

It would be interesting to use the Lax representation and the
monodromy matrix to derive finite-gap integral equations for the
deformed model analogous to those derived for strings on $\AdS$ in
\cite{KMMZ}. These equations could be then compared with the
thermodynamic limit of the Bethe equations for the deformed \N SYM
theory \cite{RR,BECH, BR}. It has been already done for the
simplest $\su(2)_\g$ case in \cite{FRT1}.

We have also discussed
string zero modes and shown that their dynamics is governed by a new fermionic generalization
of the integrable Neumann-Rosochatius model. Quantization of the model should give the
spectrum of type IIB supergravity on the $\g_i$-deformed $\AdS$ space-time.
The knowledge of the spectrum is important for checking if the nonsupersymmetric
TsT-transformed background is perturbatively stable.

The twisted boundary conditions for string coordinates may be also used to find
$1/J$ corrections to the spectrum of strings in near-plane-wave backgrounds
generalizing the computation done in \cite{Callan} for the undeformed case. The Hamiltonian
formulation, and the uniform gauge of \cite{AFu} seem to be very useful to handle the problem.
It should be also straightforward to compute the spectrum of fluctuations
around simple spinning circular strings, and analyze
$1/J$ corrections to their energies generalizing the consideration of \cite{FT2}.
In particular, it would be interesting to determine the $\g_i$-dependence of
the terms nonanalytic in $\l$ recently found in \cite{BT}, and their influence
on the dressing factor of the quantum string Bethe ansatz \cite{AFS}.

\section*{Acknowledgments}
We are grateful to A.~Tseytlin for interesting discussions. The
work of G.~A. was supported in part by the European Commission RTN
programme HPRN-CT-2000-00131, by RFBI grant N05-01-00758 and by
the INTAS contract 03-51-6346.
 The
work of S.~F.~was supported in part by the EU-RTN network {\it Constituents, Fundamental
Forces and Symmetries of the Universe} (MRTN-CT-2004-005104).
%%%%%%%%%%%%%%%%%%%%%%%%%%%%%%%%%%%%%%%%%%%%%%%%%%%%%%%%%%%%%%

\appendix

%%%%%%%%%%%%%%%%%%%%%%%%%%%%%%%%%%%%%%%%%%%%%%%%%%%%%%%%%%%%%%%%%%%%%%%%%%%%%
\section{T-duality Transformation and Rules}
%%%%%%%%%%%%%%%%%%%%%%%%%%%%%%%%%%%%%%%%%%%%%%%%%%%%%%
%\renewcommand{\theequation}{A.\arabic{equation}}
%\setcounter{equation}{0}

In this appendix we present the T-duality transformation
\cite{Buscher} for the most general Green-Schwarz action in the
form used in the paper. Our way of deriving a T-dual Green-Schwarz
action is very similar to the one used in \cite{Cvetic} where
the part of the Green-Schwarz action quadratic in fermions was also
given in an explicit form, and shown that the quadratic
fermionic term couples to background fluxes
through generalized covariant derivative.

We start with the following Green-Schwarz
action \bea \la{A1} S = -{\sqrt{\lambda}\over 2}\int\, d\tau
{d\s\over 2\pi} &&\left[ \g^{\a\b}\partial_\a \phi^i\partial_\b
\phi^j\, G_{ij}^0 - \e^{\a\b}\partial_\a \phi^i\partial_\b
\phi^j\, B_{ij}^0 \right.\\\nonumber &&\left. + 2\partial_\a
\phi^i\left( \g^{\a\b}{U}_{\b,i}^0 -\e^{\a\b}{V}_{\b,i}^0\right)
+{\cal L}_{{\rm rest}}^0\right]\, .~~~~~~~ \eea Here
$\e^{01}\equiv\e^{\tau\s}=1$ and $\g^{\a\b}\equiv \sqrt{-h}\,
h^{\a\b}$, where $h^{\a\b}$ is a world-sheet metric with Minkowski
signature. The action is invariant under U(1) isometries realized
as shifts of the angle variables $\phi_i$, $i=1,2,\ldots ,d$. We
show explicitly the dependence of the action on $\p_i$, and their
coupling to the background fields $G_{ij}^0$, $B_{ij}^0$ and
${U}_{\b,i}^0,{V}_{\b,i}^0$. These background fields are
independent of $\phi^i$ but depend on other bosonic and fermionic
string coordinates which are neutral under the U(1) isometries. By
${\cal L}_{{\rm rest}}^0$ we denote the part of the Lagrangian
which depends on these other fields of the model.

We perform a T-duality on a circle parametrized by $\p_1$. To find the T-duality rules
it is useful to represent the action (\ref{A1}) in the following equivalent form
\bea
\la{A2}
&&S = -\sqrt{\lambda}\int\, d\tau {d\s\over
2\pi} \left[ p^\a\left(\pa_\a \p_1 +{\Uh_{\a,1}^0\ov G_{11}^0}
- \g_{\a\b}\e^{\b\r}{\Vh_{\r,1}^0\ov G_{11}^0}\right)\right.
-{1\ov 2\, G_{11}^0}\g_{\a\b}\,p^\a p^\b ~~~~~~~
\\ \nonumber
&&\left. -{1\ov 2} \g^{\a\b}{\Uh_{\a,1}^0\Uh_{\b,1}^0
-\Vh_{\a,1}^0\Vh_{\b,1}^0\ov G_{11}^0}+{1\ov 2}
\e^{\a\b}{\Uh_{\a,1}^0\Vh_{\b,1}^0 -\Uh_{\b,1}^0\Vh_{\a,1}^0\ov
G_{11}^0}+{\cal L}'_{{\rm rest}} \right]\, ,
\eea
where
\bea
\Uh_{\a,1}^0\equiv U_{\a,1}^0 + \pa_\a\phi^j\, G_{1j}^0\,,\qquad
\Vh_{\a,1}^0\equiv V_{\a,1}^0 + \pa_\a\phi^j\, B_{1j}^0\,,
\eea
and
${\cal L}'_{{\rm rest}}$ denotes the part of the Lagrangian (\ref{A1})
which does not depends on $\p_1$. Indeed, varying with respect to
$p^\a$, one gets the following equation of motion for $p^\a$
\bea
\la{A3} p^\a = \g^{\a\b}\pa_\b \p_1 G_{11}^0 + \g^{\a\b}\Uh_{\b,1}^0
- \e^{\a\b}\Vh_{\b,1}^0\, .
\eea
Substituting (\ref{A3}) into
(\ref{A2}) and using the identity $\e^{\a\g}\g_{\g\r}\e^{\r\b} =
\g^{\a\b}$, we reproduce the action (\ref{A1}). Let us also
mention that up to an unessential multiplier $p^\a$ coincides with
the U(1) current corresponding to shifts of $\p_1$:
$$ p^\a \sim J_1^\a\, .$$
On the other hand, varying (\ref{A2}) with respect to $\p_1$ gives
\bea \la{A4} \pa_\a\, p^\a = 0\, . \eea The general solution to
this equation can be written in the form \bea\la{A5} p^\a =
\e^{\a\b}\pa_\b \tilde{\p}_1\, , \eea where $\tilde{\p}_1$ is the
scalar T-dual to $\p_1$. Substituting (\ref{A5}) into the action
(\ref{A2}), we obtain the following T-dual action \bea \la{A6} S =
-{\sqrt{\lambda}\over 2}\int\, d\tau {d\s\over 2\pi} &&\left[
\g^{\a\b}\partial_\a \tilde{\phi^i}\partial_\b \tilde{\phi}^j\,
\widetilde{G}_{ij} - \e^{\a\b}\partial_\a
\tilde{\phi}^i\partial_\b \tilde{\phi}^j\, \widetilde{B}_{ij}
\right.\\\nonumber &&\left. + 2\partial_\a \tilde{\phi}^i\left(
\g^{\a\b}\widetilde{U}_{\b,i}
-\e^{\a\b}\widetilde{V}_{\b,i}\right) +\widetilde{{\cal L}}_{{\rm
rest}}\right]\, .~~~~~~~ \eea with the new fields
$\widetilde{G}_{ij}$, etc  given in terms of the original ones.
\bea\la{A7} &&\widetilde{G}_{11} = {1\ov G_{11}^0}\, , \quad
\widetilde{G}_{ij} = G_{ij}^0
-{G_{1i}^0G_{1j}^0-B_{1i}^0B_{1j}^0\ov G_{11}^0}\, , \quad
\widetilde{G}_{1i}= {B_{1i}^0\ov G_{11}^0}\, ,
\\ \nonumber
&&
\widetilde{B}_{ij} = B_{ij}^0 -{G_{1i}^0B_{1j}^0-B_{1i}^0G_{1j}^0\ov G_{11}^0}\, ,
\quad \widetilde{B}_{1i}= {G_{1i}^0\ov G_{11}^0}\, ,
\\ \nonumber
&&
\widetilde{U}_{\a,1} = {V_{\a,1}^0\ov G_{11}^0}\, ,
\quad \widetilde{V}_{\a,1} = {U_{\a,1}^0\ov G_{11}^0}\, ,
\\ \nonumber
&&
\widetilde{U}_{\a,i}  = U_{\a, i}^0 -{G_{1i}^0U_{\a,1}^0-B_{1i}^0V_{\a,1}^0\ov G_{11}^0}\, ,
\\ \nonumber
&& \tilde{V}_{\a,i}  = V_{\a, i}^0
-{G_{1i}^0V_{\a,1}^0-B_{1i}^0U_{\a,1}^0\ov G_{11}^0}\, ,
\\ \nonumber
&&
{\widetilde {\cal L}}_{{\rm rest}}={\cal L}_{{\rm rest}}^0
-\g^{\a\b}{U_{\a,1}^0U_{\b,1}^0-V_{\a,1}^0V_{\b,1}^0\ov G_{11}^0} +
\e^{\a\b}{U_{\a,1}^0V_{\b,1}^0-V_{\a,1}^0U_{\b,1}^0\ov G_{11}^0}
\, ,
\\ \la{A8}
&&\e^{\a\b}\pa_\b \tilde{\p}^1= \g^{\a\b}\pa_\b \p^i G_{1i}^0 -
\e^{\a\b}\pa_\b \p^i B_{1i}^0 + \g^{\a\b}U^0_{\b,1} -
\e^{\a\b}V^0_{\b,1} \, ,\\\nonumber &&\tilde{\p}^i = \p^i\, ,\quad
i\ge 2\, . \eea In principle these formulas can be used to find
the T-duality transformed NS-NS and RR fields (see e.g. \cite{RK})
of the background in which strings propagate.

\section{The background after TsT transformation}
By using the formulas obtained in appendix A, and performing a TsT
transformation one finds the TsT-transformed background fields
$G_{ij}$, etc
\begin{eqnarray}
G_{ij}&=&\frac{G_{ij}^0}{D},\hspace{0.3in}
G_{i3}=\frac{G_{i3}^0}{D}+\hat{\gamma}\frac{B_{23}^0
G_{1i}^0-B_{13}^0G_{2i}^0+B_{12}^0G_{i3}^0}{D}\\
G_{33}&=&G_{33}^0+\frac{\hat{\gamma}+ \hat{\gamma}^2 B_{12}^0
}{D}2(B_{23}^0G_{13}^0-B_{13}^0G_{23}^0)+
\\ &~& \frac{\hat{\gamma}^2}{D}\left(G_{11}^0((B_{23}^0)^2
-(G_{23}^{0})^2)+G_{22}^0((B_{13}^0)^2 -(G_{13}^{0})^2)
+2G_{12}^0(G_{23}^0G_{13}^0-B_{13}^0B_{23}^0) \right) \nonumber
\end{eqnarray}
\begin{eqnarray}
B_{12}&=&\frac{B_{12}^0}{D}+\frac{\hat{\gamma}}{D}\left((B_{12}^0)^2-(G_{12}^0)^2+G_{11}^0G_{22}^0
\right)\\
B_{i3}&=&\frac{B_{i3}^0}{D}+\frac{\hat{\gamma}}{D} \left( B_{12}^0
B_{i3}^0- G_{13}^0G_{i2}^0+G_{23}^0G_{i1}^0 \right)\\
U_{\alpha,i}&=&\frac{U_{\alpha,i}^0}{D}+\frac{\hat{\gamma}}{D}\left(B_{12}^0
U_{\alpha,i}^0+ G_{1i}^0 V_{\alpha,2}^0 -G_{2i}^0 V_{\alpha,1}^0 \right)\\
V_{\alpha,i}&=&\frac{V_{\alpha,i}^0}{D}+\frac{\hat{\gamma}}{D}\left(B_{12}^0
V_{\alpha,i}^0+ G_{1i}^0 U_{\alpha,2}^0 -G_{2i}^0 U_{\alpha,1}^0
\right)\\
U_{\alpha,3}&=&U_{\alpha,3}^0+\frac{(\hat{\gamma}+\hat{\gamma}^2
B_{12}^0) }{D} \left( \epsilon^{ij} G_{i3}^0 V_{\alpha,j}^0
-\epsilon^{ij} B_{i3}^0 U_{\alpha,j}^0 \right)+\\
&+& \frac{\hat{\gamma}^2}{D} \left( \epsilon^{ij} U_{\alpha,i}^0
\left(G_{23}^0 G_{1j}^0-G_{13}^0 G_{2j}^0 \right) + \epsilon^{ij}
V_{\alpha,i}^0 \left(-B_{23}^0 G_{1j}^0+B_{13}^0 G_{2j}^0
\right)\right) \nonumber\\
V_{\alpha,3}&=&V_{\alpha,3}^0+\frac{(\hat{\gamma}+\hat{\gamma}^2
B_{12}^0) }{D} \left( \epsilon^{ij} G_{i3}^0 U_{\alpha,j}^0
-\epsilon^{ij} B_{i3}^0 V_{\alpha,j}^0 \right)+\\
&+& \frac{\hat{\gamma}^2}{D} \left( \epsilon^{ij} V_{\alpha,i}^0
\left(G_{23}^0 G_{1j}^0-G_{13}^0 G_{2j}^0 \right) + \epsilon^{ij}
U_{\alpha,i}^0 \left(-B_{23}^0 G_{1j}^0+B_{13}^0 G_{2j}^0
\right)\right) \nonumber\\
\L_{\rm rest}&=&\L_{\rm rest}^0+
\frac{(\hat{\gamma}+\hat{\gamma}^2 B_{12}^0) }{D}\left( 2
\epsilon^{ij} (V_{0,i}^0 V_{1,j}^0 - U_{0,i}^0 U_{1,j}^0 +
\gamma^{\alpha \beta} U_{\alpha,i}^0 V_{\beta,j}^0 )
\right)+\\
&+&\frac{\hat{\gamma}^2}{D}\left( G_{ij}^0 \epsilon^{i
\tilde{i}}\epsilon^{j \tilde{j}} \gamma^{\alpha \beta}
\left(V_{\alpha,\tilde{i}}^0 V_{\beta,\tilde{j}}^0-
U_{\alpha,\tilde{i}}^0 U_{\beta,\tilde{j}}^0\right)+G_{ij}^0
\epsilon^{i \tilde{i}}\epsilon^{j \tilde{j}} \epsilon^{\alpha
\beta} U_{\alpha,\tilde{i}}^0 V_{\beta,\tilde{j}}^0 \right)
\nonumber
\end{eqnarray}
Here the indices $i,j=1,2$ define the directions of a two-torus,
while the index 3 is singled out (in case we are dealing with more
than three indices, $3$ should be replaced by a generic index $I$
different from $1$ and $2$.) . The element $D$ is given by
$$
D=1+2\hat{\gamma}B_{12}^0+\hat{\gamma}^2(G_{11}^0G_{22}^0-(G_{12}^0)^2+
(B_{12}^0)^2)\, , ~~~~~\hat{\gamma}=\sqrt{\lambda}\gamma\, .
$$

\section{The matrices}
We choose the following parametrization of the fermionic element
\begin{equation}
\label{fermb} g(\theta,\eta)=\exp{\scriptsize \left(
\begin{array}{cccccccc}
  0 & 0 & 0 & 0 & \eta^5 & \eta^6 & \eta^7 & \eta^8 \\
  0 & 0 & 0 & 0 & \eta^1 & \eta^2 & \eta^3 & \eta^4 \\
  0 & 0 & 0 & 0 & \theta^1 & \theta^2 & \theta^3 & \theta^4 \\
  0 & 0 & 0 & 0 &  \theta^5 & \theta^6 & \theta^7 & \theta^8 \\
 \eta_5 & \eta_1 & -\theta_1 & -\theta_5  & 0 & 0 & 0 & 0 \\
 \eta_6 & \eta_2 & -\theta_2 & -\theta_6  & 0 & 0 & 0 & 0 \\
 \eta_7 & \eta_3 & -\theta_3 & -\theta_7  & 0 & 0 & 0 & 0 \\
 \eta_8 &\eta_4 &  -\theta_4 & -\theta_8  & 0 & 0 & 0 & 0
\end{array}\right)
}\, .
%\equiv\exp(Q(\theta)+S(\eta))
\end{equation}
Here $\theta^{\a}$ and $\eta^{\a}$ are $8+8$ complex fermions
obeying the following conjugation rule $\theta^{\a~*}=\theta_{\a}$
and $\eta^{\a~*}=\eta_{\a}$. Under dilatation  the fermions
$\eta^{\a}$ and $\theta^{\a}$ have charges $\sfrac{1}{2}$  and
$-\sfrac{1}{2}$ respectively \cite{Alday:2005jm}. This explains
the notational distinction we made for the fermions $\eta$'s and
$\theta$'s. In what follows it is useful to introduce the unifying
notation $\theta_{\a}$ for fermionic variables. We therefore
identify $\eta_{\a}\equiv\theta_{\a+8}$ with $\a=1,\ldots, 8$.
\medskip

In section 4 to describe the fermionic Neumann-Rosochatius model
we have used the following matrices \bea \nonumber
\begin{array}{ll}
\Upsilon_r^1=\sigma_3\otimes {\mathbb I}_2\otimes
(-i\sigma_2)\otimes \sigma_3
~~~~~~~&~~~~~~~\Upsilon_{\rho}^1=(-i\sigma_2)\otimes\sigma_1\otimes
{\mathbb I}_2\otimes {\mathbb I}_2
 \\
 \Upsilon_r^2=\sigma_3 \otimes {\mathbb I}_2\otimes (-i\sigma_2)\otimes
 \sigma_1
~~~~~~~&~~~~~~~ \Upsilon_{\rho}^2=i\sigma_2\otimes\sigma_3\otimes
{\mathbb I}_2\otimes{\mathbb I}_2 \\
 \Upsilon_r^3=\sigma_3 \otimes {\mathbb I}_2\otimes {\mathbb I}_2\otimes (-i\sigma_2)
~~~~~~~&~~~~~~~ \Upsilon_{\rho}^3=\sigma_3\otimes i\sigma_2\otimes
{\mathbb I}_2\otimes{\mathbb I}_2
\end{array}
\eea
\medskip

\noindent To present the matrices $\Omega^{ij}$ we introduce the
three auxiliary $4\times 4$ matrices $\Delta_i$: \bea \nonumber
\Delta_1=\scriptsize{\left(
\begin{array}{cccc}
 & \Lambda_1 & & \\
 & & & -\Lambda_2 \\
-\Lambda_3 & & & \\
& & \Lambda_4 &
\end{array}
\right) }\, , ~~~~ \Delta_2=\scriptsize{\left(
\begin{array}{cccc}
 &  & & \Lambda_1\\
 & & \Lambda_2 &  \\
 & -\Lambda_3 & & \\
-\Lambda_4 & &  &
\end{array}
\right) }\, ,~~~~ \Delta_3=\scriptsize{\left(
\begin{array}{cccc}
 & \Lambda_1 & & \\
-\Lambda_2 & & &  \\
 & & & \Lambda_3\\
& & -\Lambda_4 &
\end{array}
\right) }\, .
 \eea
With this definition the matrices $\Omega^{ij}$ can be written as
\bea \nonumber \Omega^{i1}=\scriptsize{\left(
\begin{array}{cccc}
 &  & & -\Delta_i\\
 & & -\Delta_i &  \\
 & \Delta_i & & \\
\Delta_i & &  &
\end{array}
\right) }\, ,~~~~ \Omega^{i2}=\scriptsize{\left(
\begin{array}{cccc}
 & & \Delta_i  & \\
 & & & -\Delta_i  \\
-\Delta_i & & & \\
& \Delta_i &  &
\end{array}
\right) }\, ,~~~~ \Omega^{i3}=\scriptsize{\left(
\begin{array}{cccc}
 & -\Delta_i & & \\
\Delta_i & & &  \\
 & & & \Delta_i\\
& & -\Delta_i &
\end{array}
\right) }\, .
 \eea
\medskip

\noindent Next we describe the structure of the matrix
$\Sigma^{ij}$ which depends on the deformation parameters
$\gamma_i$ and is symmetric under $i\leftrightarrow j$. We find
that $\Sigma$ is block-diagonal,
$\Sigma^{ij}=(-\omega^{ij},-\omega^{ij},\omega^{ij},\omega^{ij})$,
where the symmetric $4\times 4$ matrices $\omega^{ij}$ read as
\bea \nonumber \omega^{11}&=&\scriptsize{2\nu_1\left(
\begin{array}{cccc}
\nu_1+\nu_2+\nu_3 &  & & \\
 & \nu_1+\nu_2-\nu_3 & &  \\
 & & \nu_1-\nu_2-\nu_3 & \\
& &  & \nu_1-\nu_2+\nu_3
\end{array}
\right) }\, , \\ \nonumber \omega^{22}&=& \scriptsize{2\nu_2\left(
\begin{array}{cccc}
\nu_1+\nu_2+\nu_3 &  & & \\
 & \nu_1+\nu_2-\nu_3 & &  \\
 & & -\nu_1+\nu_2+\nu_3 & \\
& &  & -\nu_1+\nu_2-\nu_3
\end{array}
\right) }\, ,
\\ \nonumber \omega^{33}&=& \scriptsize{2\nu_3\left(
\begin{array}{cccc}
\nu_1+\nu_2+\nu_3 &  & & \\
 & -\nu_1-\nu_2+\nu_3 & &  \\
 & & -\nu_1+\nu_2+\nu_3 & \\
& &  & \nu_1-\nu_2+\nu_3
\end{array}
\right) }\, ,\\
\nonumber
 \omega^{12}&=& \scriptsize{\nu_3\left(
\begin{array}{cccc}
 & \nu_1-\nu_2 & & \\
\nu_1-\nu_2 &  & &  \\
 & &  & -\nu_1-\nu_2 \\
& & -\nu_1-\nu_2 &
\end{array}
\right) }\, , \\ \nonumber \omega^{13}&=& \scriptsize{\nu_2\left(
\begin{array}{cccc}
 &  & & \nu_3-\nu_1 \\
 &  & \nu_1+\nu_3 &  \\
 & \nu_1+\nu_3 &  & \\
\nu_3-\nu_1 & &  &
\end{array}
\right) }\, ,
\\ \nonumber \omega^{23}&=& \scriptsize{\nu_1\left(
\begin{array}{cccc}
 &  & \nu_2-\nu_3 & \\
 &  & & -\nu_2-\nu_3 \\
\nu_2-\nu_3 & &  & \\
& -\nu_2-\nu_3 &  &
\end{array}
\right) }\, .
 \eea

\medskip

\noindent Finally we collect the 16 by 16 matrices, $T_1$ and $T_2$:
\begin{eqnarray}
T_1&=&\left(
-\frac{S_1^2}{\rho_1^2}-\frac{S_2^2}{\rho_2^2}+\frac{S_3^2}{\rho_3^2}+\frac{J_1^2}{r_1^2}+
\frac{J_2^2}{r_2^2}+\frac{J_3^2}{r_3^2}\right)M^1 \otimes M^0 +
S_1 S_3 \left( \frac{1}{\rho_1^2}-\frac{1}{\rho_3^2} \right) M^3
\otimes
M^0  \nonumber \\
&+&S_1 S_2 \left( \frac{1}{\rho_1^2}+\frac{1}{\rho_2^2} \right)
M^0 \otimes M^0+S_1 J_1 \left( \frac{1}{r_1^2}-\frac{1}{\rho_1^2}
\right) M^2 \otimes M^2+S_1 J_2 \left(
\frac{1}{r_2^2}-\frac{1}{\rho_1^2}
\right) M^2 \otimes M^3\nonumber \\
&+&S_1 J_3 \left( \frac{1}{r_3^2}-\frac{1}{\rho_1^2} \right) M^2
\otimes M^1 +S_2 S_3 \left( \frac{1}{\rho_3^2}-\frac{1}{\rho_2^2}
\right) M^2 \otimes M^0 +S_2 J_1 \left(
\frac{1}{\rho_2^2}-\frac{1}{r_1^2} \right) M^3 \otimes M^2 \nonumber \\
&+& S_2 J_2 \left( \frac{1}{\rho_2^2}-\frac{1}{r_2^2} \right) M^3
\otimes M^3 +S_2 J_3 \left( \frac{1}{\rho_2^2}-\frac{1}{r_3^2}
\right) M^3 \otimes M^1 + S_3 J_1 \left(-
\frac{1}{\rho_3^2}-\frac{1}{r_1^2} \right) M^0 \otimes M^2 \nonumber \\
&+& S_3 J_2 \left(- \frac{1}{\rho_3^2}-\frac{1}{r_2^2} \right) M^0
\otimes M^3 + S_3 J_3 \left(- \frac{1}{\rho_3^2}-\frac{1}{r_3^2}
\right) M^0 \otimes M^1 -J_1 J_2 \left(
\frac{1}{r_1^2}+\frac{1}{r_2^2} \right) M^1 \otimes M^1 \nonumber \\
&+&J_1 J_3 \left( \frac{1}{r_1^2}+\frac{1}{r_3^2} \right) M^1
\otimes M^3+J_2 J_3 \left( \frac{1}{r_2^2}+\frac{1}{r_3^2} \right)
M^1 \otimes M^2\, , \la{T1}
\end{eqnarray}
\begin{eqnarray}\la{T2}
T_2 = G^0 \otimes M^0 +...+G^3 M^3+M^0 \otimes \tilde{G}^0+...+M^3
\otimes \tilde{G}^3\, .
\end{eqnarray}
Here the $M$'s are the following diagonal 4 by 4 matrices
\begin{eqnarray}
M^0&=&{\rm diag}(1,1,1,1),\hspace{0.4in}~M^1={\rm diag}(1,1,-1,-1),\\
M^2&=&{\rm diag}(1,-1,1,-1),\hspace{0.2in}M^3={\rm
diag}(1,-1,-1,1)\,
\end{eqnarray}
and $G$ and $\tilde{G}$ are 4 by 4, symmetric matrices, with zeros
in the diagonal. Decomposing them in terms of the following
orthogonal ``basis"
\begin{eqnarray}
\nonumber O_1&=& \scriptsize{\left(
\begin{array}{cccc}
 0& 1& 0& 0\\
 1& 0& 0& 0 \\
0 & 0& 0& 1\\
0& 0& 1 & 0
\end{array} \right)} ,\hspace{0.2in}~~~~ O_2=\scriptsize{\left(
\begin{array}{cccc}
 0& 1& 0& 0\\
 1& 0& 0& 0 \\
0 & 0& 0& -1\\
0& 0& -1 & 0
\end{array} \right)},\hspace{0.2in}~ O_3=\scriptsize{\left(
\begin{array}{cccc}
 0& 0& 1& 0\\
 0& 0& 0& 1 \\
1 & 0& 0& 0\\
0& 1& 0 & 0
\end{array} \right)}\, ,\\
O_4&=&\scriptsize{\left(
\begin{array}{cccc}
 0& 0& 1& 0\\
 0& 0& 0& -1 \\
1 & 0& 0& 0\\
0& -1& 0 & 0
\end{array} \right)},\hspace{0.2in} O_5=\scriptsize{\left(
\begin{array}{cccc}
 0& 0& 0& 1\\
 0& 0& 1& 0 \\
0 & 1& 0& 0\\
1& 0& 0 & 0
\end{array} \right)},\hspace{0.2in}~~~~~ O_6=\scriptsize{\left(
\begin{array}{cccc}
 0& 0& 0& 1\\
 0& 0& -1& 0 \\
0 & -1& 0& 0\\
1& 0& 0 & 0
\end{array} \right)}\,
\end{eqnarray}
we can show their explicit dependence on the coordinates and the
currents
\begin{eqnarray}
G^0&=&-\frac{S_2S_3\rho_1}{\rho_2
\rho_3^2}O_1-\frac{S_1S_3\rho_2}{\rho_1
\rho_3^2}O_2+\frac{S_2S_3\rho_1}{\rho_3
\rho_2^2}O_3-\frac{S_1S_2\rho_3}{\rho_1
\rho_2^2}O_4-\frac{S_1S_2\rho_3}{\rho_2
\rho_1^2}O_5-\frac{S_1S_3\rho_2}{\rho_3
\rho_1^2}O_1 \nonumber\\
G^1&=&\frac{S_1J_3\rho_2}{\rho_1
r_3^2}O_1+\frac{S_2J_3\rho_1}{\rho_2
r_3^2}O_2-\frac{S_1J_3\rho_3}{\rho_1
r_3^2}O_3+\frac{S_3J_3\rho_1}{\rho_3
r_3^2}O_4+\frac{S_3J_3\rho_2}{\rho_3
r_3^2}O_5+\frac{S_2J_3\rho_3}{\rho_2
r_3^2}O_6 \nonumber\\
G^2&=&\frac{S_1J_1\rho_2}{\rho_1
r_1^2}O_1+\frac{S_2J_1\rho_1}{\rho_2
r_1^2}O_2-\frac{S_1J_1\rho_3}{\rho_1
r_1^2}O_3+\frac{S_3J_1\rho_1}{\rho_3
r_1^2}O_4+\frac{S_3J_1\rho_2}{\rho_3
r_1^2}O_5+\frac{S_2J_1\rho_3}{\rho_2 r_1^2}O_6 \nonumber\\
G^3&=&\frac{S_1J_2\rho_2}{\rho_1
r_2^2}O_1+\frac{S_2J_2\rho_1}{\rho_2
r_2^2}O_2-\frac{S_1J_2\rho_3}{\rho_1
r_2^2}O_3+\frac{S_3J_2\rho_1}{\rho_3
r_2^2}O_4+\frac{S_3J_2\rho_2}{\rho_3
r_2^2}O_5+\frac{S_2J_2\rho_3}{\rho_2
r_2^2}O_6 \nonumber\\
\tilde{G}^0&=&-\frac{S_3J_1r_2}{r_1
\rho_3^2}O_1+\frac{S_3J_2r_1}{r_2
\rho_3^2}O_2-\frac{S_3J_2r_3}{r_2
\rho_3^2}O_3+\frac{S_3J_3r_2}{r_3
\rho_3^2}O_4-\frac{S_3J_3r_1}{r_3
\rho_3^2}O_5+\frac{S_3J_1r_3}{r_1 \rho_3^2}O_6 \nonumber\\
\tilde{G}^1&=&-\frac{J_2J_3r_1}{r_2 r_3^2}O_1+\frac{J_1J_3r_2}{r_1
r_3^2}O_2-\frac{J_1J_3r_2}{r_3 r_1^2}O_3+\frac{J_1J_2r_3}{r_2
r_1^2}O_4-\frac{J_1J_2r_3}{r_1
r_2^2}O_5+\frac{J_2J_3r_1}{r_3 r_2^2}O_6 \nonumber\\
\tilde{G}^2&=&-\frac{S_1J_1r_2}{r_1
\rho_1^2}O_1+\frac{S_1J_2r_1}{r_2
\rho_1^2}O_2-\frac{S_1J_2r_3}{r_2
\rho_1^2}O_3+\frac{S_1J_3r_1}{r_3
\rho_1^2}O_4-\frac{S_1J_3r_1}{r_3
\rho_1^2}O_5+\frac{S_1J_1r_3}{r_1
\rho_1^2}O_6 \nonumber\\
\tilde{G}^3&=&\frac{S_2J_1r_2}{r_1
\rho_2^2}O_1-\frac{S_2J_2r_1}{r_2
\rho_2^2}O_2+\frac{S_2J_2r_3}{r_2
\rho_2^2}O_3-\frac{S_2J_3r_2}{r_3
\rho_2^2}O_4+\frac{S_2J_3r_1}{r_3
\rho_2^2}O_5-\frac{S_2J_1r_3}{r_1 \rho_2^2}O_6 \nonumber
\end{eqnarray}

%%%%%%%%%%%%%%%%%%%%%%%%%%%%%%%%%%%%%%%%%%%%%%%%%%%%%%%%%%%%%%%%%%%%%%%%%%%%%

%%%%%%%%%%%%%%%%%%%%%%%%%%%%%%%%%%%%%%%%%%%%%%%%%

\end{document}